\documentclass[preprint,showpacs,superscriptaddress]{revtex4}
\usepackage{amsmath}
\usepackage{amsfonts}
\usepackage{amssymb}

\def\openone{\leavevmode\hbox{\small1\kern-3.8pt\normalsize1}}
\def\N{\leavevmode\hbox{ Z \kern-8 pt\normalsize{Z}}}
\def\openone{\leavevmode\hbox{\small1\kern-3.8pt\normalsize1}}
\def\openJ{\leavevmode\hbox{J \kern-9.5pt\normalsize J}}
\def\openS{\leavevmode\hbox{ S \kern-9.3pt\normalsize S}}
\newcommand{\bb}{\begin{equation}}
\newcommand{\ee}{\end{equation}}
\newcommand{\eqb}{\begin{eqnarray}}
\newcommand{\eqf}{\end{eqnarray}}

\usepackage{color}
\usepackage{graphicx}
\begin{document}

\title{Spinning massive test particles in cosmological and general static spherically symmetric spacetimes}

\author{Nicolas Zalaquett}
\email{nzalaque@puc.cl}
\affiliation{Facultad de F\'{\i}sica, Pontificia Universidad Cat\'olica,
Santiago, Chile.}
\author{Sergio A. Hojman}
\email{sergio.hojman@uai.cl}
\affiliation{Departamento de Ciencias,
Facultad de Artes Liberales,
Universidad Adolfo Ib\'a\~nez, Santiago, Chile.}
\affiliation{Facultad de Ingenier\'{\i}a y Ciencias,
Universidad Adolfo Ib\'a\~nez, Santiago, Chile.}
\affiliation{Departamento de F\'{\i}sica, Facultad de Ciencias, Universidad de Chile,
Santiago, Chile.}
\affiliation{Centro de Recursos Educativos Avanzados,
CREA, Santiago, Chile.}
\author{Felipe A. Asenjo}
\email{felipe.asenjo@uai.cl}
\affiliation{Facultad de Ingenier\'{\i}a y Ciencias,
Universidad Adolfo Ib\'a\~nez, Santiago, Chile.}

\begin{abstract}
A Lagrangian formalism is used to study the motion of a spinning
massive particle in Friedmann--Robertson--Walker  and  G\"odel
spacetimes, as well as in a general Schwarzschild-like spacetime and
in  static spherically symmetric conformally flat spacetimes. Exact
solutions for the motion of the particle and general exact
expressions for the momenta and velocities are displayed for
different cases. In particular, the solution for the motion in
spherically symmetric metrics is presented in the equatorial plane.
The exact solutions are found using constants of motion of the
particle, namely its mass, its spin, its angular momentum,  and a
fourth constant, which is its energy when the metric is time
independent, and a different constant otherwise. These constants are
associated to Killing vectors. In the case of the motion on the
Friedmann--Robertson--Walker metric, a new constant of motion
is found. This is the fourth constant
which generalizes previously known results obtained for spinless
particles. In the case of general Schwarzschild-like spacetimes, our
results allow for the exploration of the case of the
Reissner-Nordstrom-(Anti)de Sitter metric. Finally, for the case of
the conformally flat spacetimes, the solution is explicitly
evaluated for different metric tensors associated to a universe
filled with static perfect fluids and electromagnetic radiation. For
some combination of the values of the constants of motion the
particle trajectories may exhibit spacelike velocity vectors in
portions of the trajectories.
\end{abstract}

\pacs{04.20.Cv, 04.20.Jb, 04.40.Nr, 04.90.+e}

\maketitle

\section{Introduction}

Relativistic spinless test particles follow geodesics according to
the Equivalence Principle. On the other hand, it is known that
spinning massive test particles (tops) follow non geodesic paths
when moving on gravitational fields
\cite{mat,pap1,hojman1,hojman2,hojman2a}. The pioneering works of
Mathisson \cite{mat} and Papapetrou \cite{pap1} showed that the
equations of motion for tops are non geodesic, deriving them as
limiting cases of rotating fluids moving in gravitational fields. On
the contrary, massless spinning particles (such as photons) do
follow null geodesics as showed by Mashhoon \cite{mas} who used the
Mathisson--Papapetrou formalism for his derivation. Thus, one can
argue that the Equivalence Principle (interpreted as stating that
test particles in a gravitational field follow geodesics) is valid
only for spinless point test particles. Extended particles are, in
general, subject to tidal forces and follow, therefore, non geodesic
paths.

The rigorous derivation for the non geodesic equations of motion
obeyed by tops moving on a gravitational background can be obtained
using a Lagrangian formalism. The first derivation was obtained by
Hojman \cite{hojman1,hojman2} (using a flat spacetime formalism
developed  by Hanson and Regge \cite{hr}). In this Lagrangian
formulation, the velocity $u^\mu$ and the canonical momentum $P^\mu$
vectors are, in general, not parallel. For the motion of tops in the
electromagnetic and/or gravitational fields, the square of the mass
$m^2(\equiv P^\mu P_\mu> 0)$ is conserved implying that the momentum
vector remains timelike along the motion. However, the velocity
vector may become spacelike \cite{hojman1,vz,hr,hojman3}. It is
worth mentioning that the spin (the other Casimir function of the
Poincar\'e group) $J^2 (\equiv (1/2) S^{\mu \nu} S_{\mu \nu})$ is
also conserved for a top moving on any curved background. A proper
treatment of the lack of parallelism between velocity and momentum
is best achieved with a Lagrangian formulation of the motion of
tops, because otherwise the canonical momentum cannot be
appropriately defined. Besides, while the Mathisson--Papapetrou
formulation gives rise to third order equations of motion, the
Lagrangian approach gives rise to second order ones \cite{hojman1,
rasband}. It is important to stress that the treatment presented
here is a pole--dipole model of a particle.

Recently, the interest for the motion of tops in curved spacetimes
has staged a comeback
\cite{gane,obuk,bini,riet,hk,armenta,silenkooleg}.
 In particular, in a previous work, the motion of a top on a Schwarzschild background
was studied in detail \cite{gane}. It was found that the equations
of motion can be solved exactly, and that the spin of the particle
modifies the motion significantly as compared to the spinless
particle geodesic motion. Furthermore, this formalism has been used
to show that photons must be massless \cite{hk}.

In the same spirit, here we exhibit exact solutions for the
equations describing the motion of a top in cosmological spacetimes,
as well as in general static spherically symmetric spacetimes. The
cosmological models studied are the Friedmann--Robertson--Walker
(FRW) and the G\"odel spacetimes. The motion of a massive spinning
particle is exactly solved for motion in the equatorial plane on the
FRW metric. It is important to stress that, in this case, we find a
new conserved quantity which cannot be expressed in terms of the
Killing vectors of the metric. This constant is a generalization of
a well-known constant for spinless particles. For G\"odel spacetimes we solve completely the
motion of the spinning particle in the plane $z=0$. On the other
hand, we study a general Schwarzschild-like spacetime, showing that
it is possible to find exact solutions of the motion generalizing
the results of Ref.~\cite{gane}. This is shown for the concrete case
of the Reissner-Nordstrom-(Anti)de Sitter metric. Finally, a static
spherically symmetric conformally flat spacetime is studied. To show
the dependence of the exact solution for the orbits of the top on
the conformal factor of the metric, we explore the conformally flat
spacetime of a universe composed by radiation and a static
 perfect fluid with a cosmological equation of state. We exhibit
the exact solution for the cases of a universe filled with uniform
electromagnetic radiation only, a matter-dominated universe, a
radiation-dominated universe, and a
universe filled with dark energy.

In general,  the motion of the top in every metric is described in
terms of the constants of motion of the particle: its mass, its
spin, its angular momentum, and its energy. The latter are
constructed from the symmetries (Killing vectors) of the metric
tensor. Also, for all the models, general exact expressions for the
momenta and velocities are shown.

In addition to the general exact solutions for the motion of the
spinning massive particle, an interesting consequence of the top's
spin is highlighted. We show that, in these metrics, massive tops
described by this theory may reach spacelike velocities in portion
of the trajectories. This can be achieved if their constants of
motion satisfy certain relations. This kind of behavior is extensive
to almost all the metric studied in this work, and it seems to be a
robust effect of the motion of a spinning massive particle. Similar
results were obtained for a trajectory of a top on a Schwarzschild
spacetime \cite{gane}. This remarkable outcome is, nevertheless, not
uncommon. Theoretical results involving superluminal propagation of
massive spinning particles and fields in interaction with
electromagnetic or gravitational fields have been previously
reported in the literature by Velo and Zwanziger \cite{vz}, Hanson
and Regge \cite{hr}, Hojman \cite{hojman1} and Hojman and Regge
\cite{hojman3}. On the other hand, although some experiments have
reported hints of superluminal group velocity in optical fibers
\cite{brunner}, there is no solid experimental evidence of
superluminal neutrino propagation \cite{hirata,capa,giani,mit}.

The paper is organized as follows: in Section II we introduce the
Lagrangian theory for the motion of the top. In Section III, we
exhibit the exact solution to the equations of motion for the FRW
metric. In Section IV we show the exact result for the top's motion
in a  G\"odel spacetime. Later we study the Schwarzschild-like
metric in Section V, in conjunction with the
Reissner-Nordstrom-(Anti)de Sitter metric. In Section VI, we study
the motion of a top in a spherical symmetric conformally flat
spacetime with the cases of a universe filled with a static  perfect
fluid and radiation. Finally, in Section VII, Conclusions are
presented.

\section{Lagrangian theory for tops in gravitational fields}

The theory of a spinning massive particle in a curved spacetime was
developed  in Refs.~\cite{hojman1,hojman2} and reviewed in
\cite{gane}. In this section, we present a brief summary of the
motion of a top on a gravitational field. For a detailed
description, we refer the reader to the previously mentioned
articles.

Let us denote the position of the relativistic (spherical) top by a
four vector $x^\mu$, while its orientation is defined by an
orthonormal tetrad ${e_{(\alpha)}}^{\mu}$. A gravitational field is
described as usual in terms of the metric field $g_{\mu \nu}$
\cite{hojman1,hojman2}. The tetrad vectors satisfy $g_{\mu \nu} \
{e_{(\alpha)}}^{\mu} \ {e_{{(\beta)}}}^{\nu} \ \equiv\ \eta_{(\alpha
\beta)}$, with $\eta_{(\alpha \beta)} \ \equiv \ \ \mbox{diag}\ (+1,
-1, -1, -1)=\eta^{(\alpha \beta)}$, and have, therefore, six
independent components. The velocity vector $u^\mu$ is defined in
terms of an arbitrary parameter $\lambda$ by
\begin{equation}
u^\mu\equiv \frac{d x^\mu}{d \lambda}\, . \label{vel}
\end{equation}
The antisymmetric angular velocity tensor $\sigma^{\mu \nu}$ is
\begin{equation}
\sigma^{\mu \nu}\ \equiv \eta^{(\alpha \beta)}
{e_{(\alpha)}}^{\mu}\frac{D{e_{{(\beta)}}}}{D \lambda} ^{\nu}\ = \ -
\ \sigma^{\nu \mu}\, ,\label{sigma}
\end{equation}
where the covariant derivative $D{e_{{(\beta)}}}^{\nu}/D\lambda $ is
defined in terms of the Christoffel symbols ${\Gamma^{\nu}}_{\rho
\tau}$, as usual, by
\begin{equation}
\frac{D{e_{{(\beta)}}}^\nu}{D \lambda} \ \equiv \
\frac{d{e_{{(\beta)}}}^\nu}{d \lambda}\ + {\Gamma^{\nu}}_{\rho \tau} \
{e_{{(\beta)}}}^{\rho}\ u^\tau\, . \label{covder}
\end{equation}

The general covariance is achieved unambiguously at the level of the
Lagrangian formulation \cite{hojman1} due to the fact that only
first derivatives of the dynamical variables are used in its
construction. If no Lagrangian theory for a system of special
relativistic equations of motion is known, the introduction of
gravitational interactions cannot be unambiguously implemented.

A possible Lagrangian $L = L (a_1, a_2, a_3, a_4)$ is constructed as
an arbitrary function of four invariants $a_1\equiv u^\mu u_\mu,\
a_2\equiv \sigma^{\mu \nu} \sigma_{\mu \nu}= -
\mbox{tr}({\sigma}^2),\ a_3\equiv u_\alpha \sigma^{\alpha \beta}
\sigma_{\beta \gamma} u^\gamma$ and $a_4\equiv
\mbox{det}({\sigma})$,
\begin{equation}
L (a_1, a_2, a_3, a_4) = (a_1)^{1/2}\mathcal{L} \left(a_2/a_1,
a_3/(a_1)^2, a_4/(a_1)^2\right)\, , \label{lag1}
\end{equation}
such that the action $S=\int L\, d\lambda$, be
$\lambda$--reparametrization invariant (the speed of light $c$ is
set equal to $1$). $\mathcal{L}$ is an arbitrary function of three
variables. Note that, unlike the spinless case, it is not necessary
that $a_1$ be positive to have a real Lagrangian (see
Ref.~\cite{gane} for an extended discussion about this issue).

The conjugated momentum vector $P_\mu$ and antisymmetric spin tensor
$S_{\mu \nu}$ are defined by
\begin{equation}
P_\mu \equiv \frac{\partial L}{\partial u^\mu}\, ,\qquad
S_{\mu \nu} \equiv \frac{\partial L}{\partial \sigma^{\mu \nu}} = -
S_{\nu \mu}\, .\label{sigmamunu}
\end{equation}

As usual, the equations of motion are obtained by considering the
variation of the action $S$ with respect to (ten) independent
variations $\delta x^\mu$ and (the covariant generalization of)
$\delta \theta^{\mu \nu} \equiv \eta^{(\alpha \beta)}
{e_{(\alpha)}}^{\mu}{\delta {e_{{(\beta)}}}}^\nu\ = - \delta
\theta^{\nu \mu}$. The final equations of motion turn out to be non
geodesic \cite{hojman1, hojman2}
\begin{equation}
 \frac{D P^\mu}{D\lambda}=-\frac{1}{2}{R^\mu}_{\nu\alpha\beta}u^\nu S^{\alpha\beta}\, ,
\label{momentummotion}
\end{equation}
and
\begin{equation}
\frac{D S^{\mu \nu}}{D\lambda}=S^{\mu
\lambda}{\sigma_\lambda}^\nu-\sigma^{\mu
\lambda}{S_\lambda}^\nu=P^\mu u^\nu-u^\mu P^\nu\, . \label{spinmotion}
\end{equation}

These results hold for arbitrary $\cal L$. The dynamical variables
$P^\mu$ and $S^{\mu \nu}$ are interpreted as the ten generators of
the Poincar\'e group.

In order to restrict the spin tensor to generate rotations only, the
Tulczyjew constraint \cite{tulc} is considered \cite{hojman1, hr}
\begin{equation}
S^{\mu \nu} P_\nu=0\, .\label{constraint}
\end{equation}
However, we would like to stress that, for a  suitably chosen  $\cal
L$, this constraint is a consequence of the theory, i.e., the
Tulczyjew constraint is derived from this Lagrangian formalism (for
details, see Ref.~\cite{gane}). This is one of the strengths of this
Lagrangian theory.

It can also be proved \cite{gane} that both the top mass $m$ and its
spin $J$ are conserved quantities
\begin{equation}
m^2\equiv P^\mu P_\mu\, , \label{mass}
\end{equation}
\begin{equation}
J^2\equiv \frac{1}{2} S^{\mu \nu}S_{\mu \nu}\, . \label{spin}
\end{equation}

Finally, a conserved quantity $C_{\xi}$ given by
\begin{equation}
C_{\xi}\equiv P^\mu \xi_\mu -\frac{1}{2}S^{\mu \nu}\xi_{\mu;\nu} \,
, \label{ckilling}
\end{equation}
can be associated to any Killing vector $\xi_\mu$ of the metric
\begin{equation}
\xi_{\mu;\nu} + \xi_{\nu;\mu}=0 \, . \label{killingeq}
\end{equation}

As the general theory is established, we proceed now in the
following sections to find the exact solutions for the motion of a
top in cosmological spacetimes using the
Friedmann--Robertson--Walker and the G\"odel metrics. We also study
the motion in general static spherically symmetric spacetimes with
Schwarzschild-like and conformally flat metrics.

\section{Exact solution for Cosmological Friedmann--Robertson--Walker spacetimes}

Let us consider as a first case the Friedmann--Robertson--Walker
(FRW) metric, which is given by the following line element in
spherical coordinates
\begin{equation}\label{metriConfFRW}
ds^2=g_{\mu\nu}d x^\mu d x^\nu=c^2 dt^2-a(t)^2 g(r) dr^2-a(t)^2 r^2 (d\theta^2+\sin^2\theta d\phi^2)\, ,
\end{equation}
where $r$ is the radial distance, $\theta$ and $\phi$ are the polar
and azimuthal angles respectively, and $g(r)=1/(1-kr^2)$. Here
$a(t)\equiv a$ is the time-dependent scale factor of the universe,
whereas $k$ assumes three possible values $k=-1,0,1$, denoting a
universe with negative spatial curvature, spatially flat or with
positive spatial curvature respectively. From now on, we explicitly
display the speed of light $c$  in our calculations.

Before solving the equations of motion for the top, it is useful to
exhibit the three Killing vectors of the FRW metric explicitly
\begin{equation}
\begin{aligned}
 &\xi^{0}_\mu=(0,0,0,-a^2 r^2 \sin^2\theta)\, , \\
 &\xi^{1}_\mu=(0,0,a^2 r^2 \sin\phi, a^2 r^2 \sin\theta\cos\theta\cos\phi)\, ,\\
 &\xi^{2}_\mu=(0,0,-a^2 r^2 \cos\phi, a^2 r^2 \sin\theta\cos\theta\sin\phi)\, . \\
\end{aligned}
\end{equation}

Using these Killing vectors, we can solve the equations of motion
\eqref{momentummotion} and \eqref{spinmotion} in general. In what
follows, we study the motion in the equatorial plane $\theta=\pi/2$,
which is defined to be orthogonal to the conserved angular momentum
vector. This choice simplifies the analysis because, in this case,
$S^{r\theta}=S^{t\theta}=S^{\phi\theta}=0$, and also
$\dot\theta=0=P^\theta=\dot P^\theta$. Thus, the top remains in the
plane orthogonal to the total angular momentum if it was initially
there.

We will make use of constants of motion to find the top's trajectory
in a FRW spacetime. The mass $m^2$ and spin $J^2$ are always
conserved in any spacetime metric. The formalism provides three
constants of motion, which can be found using the three (angular
momentum) Killing vectors. Two of the three components of the
angular momentum vector have been used to define the equatorial
plane. The third component $j$ is the angular momentum component
perpendicular to the equatorial plane. The problem can be completely
solved if we find a fourth constant of motion. In this case, the
energy of the top motion is not conserved because the metric is
time-dependent. Therefore, to
find the last constant we have to integrate the equations of motion.

From the constants of motion \eqref{mass} and \eqref{spin} we find
\begin{eqnarray}
m^2 c^2=c^2 (P^t)^2-a^2 \left[g (P^r)^2+r^2 (P^\phi)^2\right]\, , \label{constrMFRW}\\
J^2=a^2 g \left(a^2 r^2 \left({S}^{r\phi}\right)^2-c^2 \left({S}^{{tr}}\right)^2\right)-a^2c^2 r^2 \left({S}^{{t\phi}}\right)^2\, , \label{constrJFRW}
\end{eqnarray}
where we can identify $J$ with the top's spin. Making use of the
Killing vectors, we find another constant which corresponds to the
conserved angular momentum component orthogonal to the orbital plane
\begin{equation}\label{jmomFRW}
j=-a r \left[r \dot a {S}^{{t\phi}}+a \left(r {P}^{\phi}+{S}^{{r\phi}}\right)\right]\, ,
\end{equation}
where $\dot a\equiv d a/dt$.

To find the fourth constant, we first display the Tulczyjew
contraints \eqref{constraint}
\begin{eqnarray}
a^2 r^2 {P}^{\phi} {S}^{{t\phi}}+a^2 g {P}^{{r}} {S}^{{tr}}=0\, , \label{cT1}\\
c^2 {P}^{{t}} {S}^{{tr}}+a^2 r^2 {P}^{\phi} {S}^{{r\phi}}=0\, , \label{cT2}\\
 a^2 g {P}^{{r}} {S}^{{r\phi}}-c^2 {P}^{{t}} {S}^{{t\phi}}=0\, . \label{cT3}
\end{eqnarray}

The explicit form of the equations of motion can be obtained form
\eqref{momentummotion} and \eqref{spinmotion}. The equations of
motion for the momentum \eqref{momentummotion} are
\begin{eqnarray}
\frac{a g \dot{r} \ddot{a} {S}^{{tr}}}{c^2}+\frac{a r^2 \dot{\phi } \ddot{a} {S}^{{t\phi}}}{c^2}+\frac{a g \dot{r} \dot{a} {P}^{{r}}}{c^2}+\frac{a r^2 \dot{\phi } \dot{a} {P}^{\phi }}{c^2}+\dot{{P}^{{t}}}=0\, , \label{momentfirstC}\\
 \frac{\ddot{a} {S}^{{tr}}}{a}+\frac{r^2 \dot{\phi } \dot{a}^2 {S}^{{r\phi}}}{c^2}+\frac{\dot{a} {P}^{{r}}}{a}+\frac{\dot{r} \dot{a} {P}^{{t}}}{a}+\frac{\dot{r} g' {P}^{{r}}}{2 g}+\frac{r \dot{\phi } g' {S}^{{r\phi}}}{2 g^2}-\frac{r \dot{\phi } {P}^{\phi }}{g}+\dot{{P}^{{r}}} =0\, ,\\
 \frac{\ddot{a} {S}^{{t\phi}}}{a}-\frac{g \dot{r} \dot{a}^2 {S}^{{r\phi}}}{c^2}+\frac{\dot{\phi } \dot{a} {P}^{{t}}}{a}+\frac{\dot{a} {P}^{\phi }}{a}-\frac{\dot{r} g' {S}^{{r\phi}}}{2 g r}+\frac{\dot{\phi } {P}^{{r}}}{r}+\frac{\dot{r} {P}^{\phi }}{r}+\dot{{P}^{\phi }}=0\, ,
\end{eqnarray}
where the symbol $'$ denotes derivatives with respect to $r$. On the
other hand, the equations of motion for the spin \eqref{spinmotion}
are
\begin{eqnarray}
 -\frac{a r^2 \dot{\phi } \dot{a} {S}^{{r\phi}}}{c^2}+\frac{\dot{a} {S}^{{tr}}}{a}+\frac{\dot{r} g' {S}^{{tr}}}{2 g}-\frac{r \dot{\phi } {S}^{{t\phi}}}{g}+{P}^{{r}}-\dot{r} {P}^{{t}}+\dot{{S}^{{tr}}}=0\, , \\
 \frac{a g \dot{r} \dot{a} {S}^{{r\phi}}}{c^2}+\frac{\dot{a} {S}^{{t\phi}}}{a}-\dot{\phi } {P}^{{t}}+{P}^{\phi }+\frac{\dot{\phi } {S}^{{tr}}}{r}+\frac{\dot{r} {S}^{{t\phi}}}{r}+\dot{{S}^{{t\phi}}} =0\, ,\\
\frac{\dot{r} \dot{a} {S}^{{t\phi}}}{a}+\frac{2 \dot{a} {S}^{{r\phi}}}{a}-\frac{\dot{\phi } \dot{a} {S}^{{tr}}}{a}+\frac{\dot{r} g' {S}^{{r\phi}}}{2 g}-\dot{\phi } {P}^{{r}}+\dot{r} {P}^{\phi }+\frac{\dot{r} {S}^{{r\phi}}}{r}+\dot{{S}^{{r\phi}}}=0\, .\label{spinlastS}
 \end{eqnarray}

Following with our analysis we have to use the constraints
\eqref{cT1} and \eqref{cT2}, as well as the equations for the
constants of motion \eqref{constrMFRW} and \eqref{constrJFRW}, to
get the relation
 \begin{equation}\label{SkappaPphi}
{S}^{{tr}}= \kappa {P}^{\phi }\, ,
 \end{equation}
with the function $\kappa$ defined as
 $$
 \kappa \equiv \pm\frac{J r}{c^2 \sqrt{g} m}\, .
 $$
Thus, solving for the spin components in terms of the momentum
components, and  using both constraints as well as
Eq.~\eqref{SkappaPphi}, we get
 \begin{eqnarray}
{S}^{{t\phi}}=-\frac{g \kappa  {P}^{{r}}}{r^2}\, ,\label{StphiSrphi2FRW1}\\
 {S}^{{r\phi}}=-\frac{c^2 \kappa  {P}^{{t}}}{a^2 r^2}\, ,\label{StphiSrphi2FRW2}
 \end{eqnarray}
which their time derivatives are readily calculated as
\begin{eqnarray}
\dot{{S}^{{tr}}}=\kappa  \left(\frac{(2-g) \dot{r} {P}^{\phi }}{r}+\dot{{P}^{\phi }}\right)\, ,\\
\dot{{S}^{{t\phi}}}=-\frac{g \kappa  \left((g-2) \dot{r} {P}^{{r}}+r \dot{{P}^{{r}}}\right)}{r^3}\, ,\\
\dot{{S}^{{r\phi}}}=\frac{c^2 \kappa  \left({P}^{{t}} \left(2 r \dot{a}+a g \dot{r}\right)-a r \dot{{P}^{{t}}}\right)}{a^3 r^3}\, ,
 \end{eqnarray}
where we have used the relation $g' = 2 (g-1) g/{r}$. Replacing the
spin components and their derivatives in the previous set
\eqref{momentfirstC}-\eqref{spinlastS} we can find a set of
equations for the momentum components
\begin{eqnarray}
\kappa  \left(\frac{\dot{\phi } \dot{a} {P}^{{t}}}{a}+\frac{\dot{a} {P}^{\phi }}{a}+\frac{\dot{\phi }{P}^{{r}}}{r}+\frac{\dot{r} {P}^{\phi }}{r}+\dot{{P}^{\phi }}\right)+{P}^{{r}}-\dot{r} {P}^{{t}}=0\, ,\label{eMoMe61}\\
\kappa  \left(-\frac{g \dot{a} {P}^{{r}}}{a r^2}-\frac{g \dot{r} \dot{a} {P}^{{t}}}{a r^2}-\frac{g^2 \dot{r} {P}^{{r}}}{r^3}+\frac{g \dot{r} {P}^{{r}}}{r^3}-\frac{g \dot{{P}^{{r}}}}{r^2}+\frac{\dot{\phi } {P}^{\phi }}{r}\right)-\dot{\phi } {P}^{{t}}+{P}^{\phi }=0\, ,\label{eMoMe62}\\
 \kappa  \left(-\frac{c^2 \dot{{P}^{{t}}}}{a^2 r^2}-\frac{g \dot{r} \dot{a} {P}^{{r}}}{a r^2}-\frac{\dot{\phi } \dot{a} {P}^{\phi }}{a}\right)-\dot{\phi } {P}^{{r}}+\dot{r} {P}^{\phi }=0\, ,\label{eMoMe63}\\
 \kappa  \left(\frac{a g \dot{r} \ddot{a}{P}^{\phi }}{c^2}-\frac{a g \dot{\phi } \ddot{a} {P}^{{r}}}{c^2}\right)+\frac{a g \dot{r} \dot{a} {P}^{{r}}}{c^2}+\frac{a r^2 \dot{\phi } \dot{a} {P}^{\phi }}{c^2}+\dot{{P}^{{t}}}=0\, ,\label{eMoMe64}\\
 \frac{\dot{a} {P}^{{r}}}{a}+\frac{\dot{r} \dot{a} {P}^{{t}}}{a}+\kappa  \left(\frac{c^2 \dot{\phi } {P}^{{t}}}{a^2 g r^2}-\frac{c^2 \dot{\phi } {P}^{{t}}}{a^2 r^2}+\frac{\ddot{a} {P}^{\phi }}{a}-\frac{\dot{\phi } \dot{a}^2 {P}^{{t}}}{a^2}\right)+\frac{g \dot{r} {P}^{{r}}}{r}-\frac{r \dot{\phi } {P}^{\phi}}{g}-\frac{\dot{r} {P}^{{r}}}{r}+\dot{{P}^{{r}}}=0\, ,\label{eMoMe65}\\
 \frac{\dot{\phi } \dot{a} {P}^{{t}}}{a}+\frac{\dot{a} {P}^{\phi }}{a}+\kappa  \left(\frac{c^2 g \dot{r} {P}^{{t}}}{a^2 r^4}-\frac{c^2 \dot{r} {P}^{{t}}}{a^2 r^4}-\frac{g \ddot{a} {P}^{{r}}}{a r^2}+\frac{g \dot{r} \dot{a}^2 {P}^{{t}}}{a^2 r^2}\right)+\frac{\dot{\phi } {P}^{{r}}}{r}+\frac{\dot{r} {P}^{\phi }}{r}+\dot{{P}^{\phi }}=0\, .\label{eMoMe66}
 \end{eqnarray}

The process of finding the new constant of motion consists first on
substracting Eq.~\eqref{eMoMe66} multiplied by $\kappa$ from
Eq.~\eqref{eMoMe61} to obtain
 \begin{eqnarray}
 {P}^{{r}} \left(\frac{g \kappa ^2 \ddot{a}}{a r^2}+1\right)+\dot{r} {P}^{{t}} \left(-\frac{c^2 g \kappa^2}{a^2 r^4}+\frac{c^2 \kappa^2}{a^2 r^4}-\frac{g \kappa^2 \dot{a}^2}{a^2 r^2}-1\right)=0\, .
\end{eqnarray}
In the same fashion we can add Eq.~\eqref{eMoMe62} to
Eq.~\eqref{eMoMe65} multiplied by ${\kappa g}/{r^2}$ to get
\begin{eqnarray}
{P}^{\phi} \left(\frac{g \kappa ^2 \ddot{a}}{a r^2}+1\right)+\dot{\phi} {P}^{{t}} \left(-\frac{c^2 g \kappa^2}{a^2 r^4}+\frac{c^2 \kappa^2}{a^2 r^4}-\frac{g \kappa^2 \dot{a}^2}{a^2 r^2}-1\right)=0\, .
 \end{eqnarray}
These two above equation can be solved for $P^r$ and $P^{\phi}$ in terms of $P^t$ as
 \begin{eqnarray}
 P^r=\alpha \dot{r} P^t\, , \qquad P^{\phi}=\alpha \dot{\phi} P^t\, ,\label{PryphiFRW}
 \end{eqnarray}
where the function $\alpha$ depends on time only
 $$
 \alpha \equiv \frac{a^2 c^4 m^2+J^2 \dot{a}^2+c^2 J^2 k}{a^2 c^4 m^2+a J^2 \ddot{a}}\, .
 $$
 Note that $\alpha=1$ if $J=0$.
Now, to get $P^t$ use Eqs.~\eqref{PryphiFRW} in \eqref{eMoMe63} to
find
\begin{equation}\label{72fff}
\frac{a \alpha  c^2 \dot{{P}^{{t}}} {P}^{{t}}}{\dot{a}}=\left({P}^{{t}}\right)^2 \left(-a^2 \alpha ^2 g \dot{r}^2-a^2 \alpha^2 r^2 \dot{\phi}^2\right)\, .
 \end{equation}
At the same time, using \eqref{PryphiFRW} in the constant of motion \eqref{mass} we get
 \begin{equation}\label{73fff}
 c^2 m^2=\left({P}^{{t}}\right)^2 \left(-a^2 \alpha^2 g \dot{r}^2-a^2 \alpha^2 r^2 \dot{\phi}^2+c^2\right)\, .
 \end{equation}
Subtracting \eqref{73fff} from \eqref{72fff} we finally find
 \begin{equation}
 \frac{\dot{a} \left(\left({P}^{{t}}\right)^2-m^2\right)}{a \alpha }+\dot{{P}^{{t}}} {P}^{{t}}=0\, ,
 \end{equation}
which has the solution
 \begin{equation}
 {P}^{{t}}=\sqrt{\frac{2 \sigma}{a^2 c^4 m^2+J^2 \dot{a}^2+c^2 J^2 k}+m^2}\, ,
 \end{equation}
 where $\sigma$ is a new (integration) constant of motion. This constant
 is one of the important results of this work and deserves more attention.
 The new constant can be written in terms of $P^t$ as
 \begin{eqnarray}\label{newconstantS}
 \sigma=\frac{m^2c^4}{2}\left[a^2\left(1+\frac{J^2 H^2}{m^2c^4}\right)+\frac{J^2k}{m^2c^2}\right]\left[\left(P^{t}\right)^2-m^2\right]=\frac{\beta}{2}\left[\left(P^{t}\right)^2-m^2\right]\, ,
 \end{eqnarray}
 where $H\equiv\dot a/a$ is the Hubble parameter and
$$\beta \equiv a^2  m^2 c^4+J^2 \dot{a}^2+c^2 J^2 k\, .$$
The constant \eqref{newconstantS} can be written as
$\sigma=K_{\mu\nu}P^\mu P^\nu$, associated to the tensor
\begin{equation}\label{killingTensor1}
K_{\mu\nu}=\frac{\beta}{2}\left({\cal U}_\mu {\cal U}_\nu-\frac{1}{c^2}g_{\mu\nu}\right)\, ,
\end{equation}
where ${\cal U}_\mu=(1,0,0,0)$.
 The constant $\sigma$ has a well-known analogue when $J=0$
for spinless particles \cite{Carroll8}. It is important to emphasize that $K_{\mu \nu}/m^2$ is a Killing tensor for $J=0$ only. When $J \neq 0$ we have not been able to write the new constant of motion in terms of a Killing tensor. Therefore, we have found a
generalization for that constant in the case of a massive spinning particle  $J\neq 0$ in FRW
spacetimes.
 As far as we know, this is the first time that the constant \eqref{newconstantS} has been found for the motion
 of massive spinning particles on a FRW metric.
 The new extra terms (proportional to $J^2$) are interesting.
 Note how the first of these two terms combines to introduce a
 correction of the spin that takes in account the expansion of the universe
 through the Hubble parameter. Similarly, the second term of these two terms tell
 us how the curvature of the space affects the value of this constant. As we can see,
 this constant is richer in information than the spinless case counterpart.

 Finally, making use of \eqref{newconstantS} to complete determine $P^t$, we can obtain the other momenta \eqref{PryphiFRW}.
  Also, we can calculate the line element \eqref{metriConfFRW} as
 \begin{equation}
 \frac{ds^{2}}{c^2dt^2}=1-\frac{1}{\alpha^2}-\frac{m^2}{\alpha ^2\left({P}^{{t}}\right)^2}=1-\frac{a^2 \sigma \dot{\beta}^2}{2 \beta ^2 \dot{a}^2 \left(2 \sigma+\beta  m^2\right)}\, ,
\end{equation}
which can be written in terms of the constants of motion
\begin{equation}\label{dsFRWsuperl}
\frac{ds^{2}}{c^2dt^2}=1-\frac{2 \sigma \left(J^2\dot{H}+J^2H^2+c^4 m^2\right)^2}{\left({J^2 H^2}+c^4 m^2\right)^2 \left(a^2 m^2 \left(J^2 H^2+c^4 m^2\right)+2 \sigma\right)}\, .
\end{equation}

The line element \eqref{dsFRWsuperl} for a top in FRW metric is no
longer positive definite. It can either vanish or be negative
depending on the values of the different constants involved in its
expression. Therefore, the top may follow lightlike or spacelike
trajectories when $ds^2=0$ or  $ds^2< 0$, for appropriate values of
the constants. This results are possible only if $\dot H\neq 0$.

This feature the line element on portion of the trajectories is a
characteristic of dynamics of massive spinning particles. It has
been previously reported for the top's motion in Schwarzschild
spacetimes \cite{gane}.

As a final exercise for the FRW metric, we can calculate $\dot{r}$
and $\dot{\phi}$ using the Eqs.~\eqref{jmomFRW},
\eqref{StphiSrphi2FRW1}, \eqref{StphiSrphi2FRW2} and \eqref{73fff},
to obtain
\begin{equation}
\dot{r}=\frac{\dot{\beta} \left(\sqrt{g} \kappa  \dot{a} \left(j r-c^2 \kappa  {P}^{{t}}\right)\pm a r \Upsilon\right)}{2 \beta  \sqrt{g} r \dot{a} {P}^{{t}} \left(a^2 r^2+g \kappa ^2 \dot{a}^2\right)}\, ,\qquad
\dot{\phi }=\frac{\left(\alpha  g \kappa  r \dot{r}a \dot{a} +c^2 \kappa\right)  {P}^{{t}}-j r}{a^2 \alpha  r^3 {P}^{{t}}}\, ,
\end{equation}
where
$$
\Upsilon=\sqrt{-r^2 \left(a^2 c^2 m^2 r^2+c^2 g \kappa ^2 m^2 \dot{a}^2+j^2\right)+c^2 \left({P}^{{t}}\right)^2 \left(a^2 r^4+g \kappa ^2 r^2 \dot{a}^2-c^2 \kappa ^2\right)+2 c^2 j \kappa  r {P}^{{t}}}\, .
$$

\section{Exact solution for Cosmological G\"odel spacetimes}

Consider  a universe described by the G\"odel metric which is given by \cite{HawkingEllis}
\begin{equation}\label{godelmetricg}
{g}^{\mu\nu}=\left(
\begin{array}{cccc}
 c^2 & 0 & c e^{x w_0} & 0 \\
 0 & -1 & 0 & 0 \\
 c e^{x w_0} & 0 & \frac{1}{2} e^{2 x w_0} & 0 \\
 0 & 0 & 0 & -1 \\
\end{array}
\right)\, ,
\end{equation}
in rectangular coordinates, where $w_0$ is a constant related to the angular velocity of the rotating universe.
This metric has five Killing vectors
\begin{equation}
\begin{aligned}
 &\xi^{0}_\mu=\left(c^2,0,c e^{w_0 x},0\right)\, , \\
 &\xi^{1}_\mu=\left(0, 0, 0, -1\right)\, , \\
 &\xi^{2}_\mu=\left(c e^{w_0 x},0,\frac{1}{2} e^{2 w_0 x},0\right)\, ,\\
 &\xi^{3}_\mu=\left(-c w_0 y e^{w_0 x},-1,-\frac{1}{2} w_0 y e^{2 w_0 x},0\right)\, , \\
 &\xi^{4}_\mu=\left(-\frac{1}{2} c w_0 y^2 e^{w_0 x}-\frac{c e^{-w_0 x}}{w_0},-y,-\frac{1}{4} w_0 y^2 e^{2 w_0 x}-\frac{3}{2 w_0},0\right)\, .
\end{aligned}
\end{equation}
It becomes apparent that there are solutions describing trajectories
in the plane $z=0$. Therefore, we work in this plane with $P^{z}=0$
and $\dot{P^{z}}=0$ (this also implies that every $z$-component of
the spin as well as their time derivatives vanish).

We follow the same procedure than in the previous section. First we find the constants of motion \eqref{mass} and \eqref{spin} as
\begin{eqnarray}
m^2 c^2&=&c^2 \left({P}^{{t}}\right)^2+2 c e^{w_0 x} {P}^{{t}} {P}^{{y}}+\frac{1}{2} e^{2 w_0 x} \left({P}^{{y}}\right)^2-\left({P}^{{x}}\right)^2\, , \label{GodM2}\\
J^2&=&-c^2 \left({S}^{{tx}}\right)^2-\frac{1}{2} c^2 e^{2 w_0 x} \left({S}^{{ty}}\right)^2+2 c e^{w_0 x} {S}^{{tx}} {S}^{{xy}}-\frac{1}{2} e^{2 w_0 x} \left({S}^{{xy}}\right)^2\, , \label{GodJ2}
\end{eqnarray}
while using the Killing vectors we can calculate the following non
vanishing constants of motion
\begin{eqnarray}
E&=&\frac{1}{2} c \left(2 c {P}^{{t}}+e^{w_0 x} \left(2 {P}^{{y}}+w_0 {S}^{{xy}}\right)\right)\, , \label{consEGodel}\\
C_{2}&=&\frac{1}{2} e^{w_0 x} \left(2 c {P}^{{t}}+w_0 \left(e^{w_0 x} {S}^{{xy}}-c {S}^{{tx}}\right)+e^{w_0 x} {P}^{{y}}\right)\, , \\
C_{3}&=&\frac{1}{2} \left(-w_0 e^{w_0 x} \left(2 c y {P}^{{t}}-c {S}^{{ty}}+y e^{w_0 x} {P}^{{y}}\right)+w_0^2 y \left(-e^{w_0 x}\right) \left(e^{w_0 x} {S}^{{xy}}-c {S}^{{tx}}\right)-2 {P}^{{x}}\right)\, , \\
C_{4}&=&-\frac{1}{2} c w_0 y^2 e^{w_0 x} {P}^{{t}}-\frac{c e^{-w_0 x} {P}^{{t}}}{w_0}+\frac{1}{4} c w_0^2 y^2 e^{w_0 x}{S}^{{tx}}-\frac{1}{2} c e^{-w_0 x} {S}^{{tx}}+\frac{1}{2} c w_0 y e^{w_0 x} {S}^{{ty}}\nonumber\\
&&-\frac{1}{4} w_0 y^2 e^{2 w_0 x}+ {P}^{{y}}-\frac{3 {P}^{{y}}}{2 w_0}-y {P}^{{x}}-\frac{1}{4} w_0^2 y^2 e^{2 w_0 x}{S}^{{xy}}+\frac{{S}^{{xy}}}{2}\, .
\end{eqnarray}
On the other hand, the two independent Tulczyjew constraint
equations \eqref{constraint} read
\begin{eqnarray}
{P}^{{x}} {S}^{{tx}}&=&c e^{w_0 x} {P}^{{t}} {S}^{{ty}}+\frac{1}{2} e^{2 w_0 x} {P}^{{y}} {S}^{{ty}}\, , \label{GodTul1}\\
{P}^{{x}} {S}^{{xy}}&=&c^2 {P}^{{t}} {S}^{{ty}}+c e^{w_0 x} {P}^{{y}} {S}^{{ty}}\, .\label{GodTul2}
\end{eqnarray}

The above constants of motion must be used along with the equations
of motion for the top in the G\"odel spacetimes. The (non
identically vanishing) momentum equations of motion
\eqref{momentummotion} are
\begin{eqnarray}
&&w_0 \left(2 \dot{x} \left(2 c {P}^{{t}}-c w_0 {S}^{{tx}}+e^{w_0 x} {P}^{{y}}+2 w_0 e^{w_0 x} {S}^{{xy}}\right)+\left(2 c+\dot{y} e^{w_0 x}\right) \left(c w_0 e^{w_0 x} {S}^{{ty}}+2 {P}^{{x}}\right)\right)\nonumber\\
&&\qquad\qquad\qquad\qquad\qquad\qquad\qquad\qquad\qquad\qquad\qquad\qquad\qquad\qquad\qquad\qquad+4 c \dot{{P}^{{t}}} =0\, ,\label{GodP1}\\
&&w_0 \left(\dot{y} e^{w_0 x} \left(2 c {P}^{{t}}+w_0 \left(3 e^{w_0 x} {S}^{{xy}}-2 c {S}^{{tx}}\right)+2 e^{w_0 x} {P}^{{y}}\right)+2 c \left(w_0 \left(e^{w_0 x} {S}^{{xy}}-c {S}^{{tx}}\right)+e^{w_0 x} {P}^{{y}}\right)\right)\nonumber\\
&&\qquad\qquad\qquad\qquad\qquad\qquad\qquad\qquad\qquad\qquad\qquad\qquad\qquad\qquad\qquad\qquad+4 \dot{{P}^{{x}}}=0\, ,\label{GodP2} \\
&&2 e^{w_0 x} \dot{{P}^{{y}}}-w_0 \left(\dot{x} \left(2 c {P}^{{t}}+w_0 e^{w_0 x} {S}^{{xy}}\right)+c \left(w_0 e^{w_0 x} {S}^{{ty}} \left(c+\dot{y} e^{w_0 x}\right)+2 {P}^{{x}}\right)\right)=0\, ,\label{GodP3}
\end{eqnarray}
whereas the equations for the spin \eqref{spinmotion} become
\begin{eqnarray}
 \frac{1}{2} c w_0 e^{w_0 x} {S}^{{ty}}-\frac{w_0 \dot{x} e^{w_0 x} {S}^{{xy}}}{2 c}-\dot{x} {P}^{{t}}+{P}^{{x}}+w_0 \dot{x} {S}^{{tx}}+\dot{{S}^{{tx}}}+\frac{1}{2} w_0 \dot{y} e^{2 w_0 x} {S}^{{ty}}=0\, ,\label{GodS1} \\
-c w_0 e^{-w_0 x} {S}^{{tx}}+\frac{w_0 \dot{y} e^{w_0 x} {S}^{{xy}}}{2 c}-\dot{y} {P}^{{t}}+{P}^{{y}}+w_0 \dot{x} {S}^{{ty}}+\dot{{S}^{\text{ty}}}+w_0 {S}^{{xy}} =0\, ,\label{GodS2}\\
 c w_0 \dot{x} e^{-w_0 x} {S}^{{tx}}+\frac{1}{2} c w_0 \dot{y} e^{w_0 x} {S}^{{ty}}-\dot{y} {P}^{{x}}+\dot{x} {P}^{{y}}+\dot{{S}^{{xy}}}=0\, .\label{GodS3}
 \end{eqnarray}

Now, using Eqs.~\eqref{GodM2}, \eqref{GodJ2},  \eqref{GodTul1}, and \eqref{GodTul2}, we can find the first solution for the spin component
 \begin{equation}
 {S}^{{ty}}=e^{-w_0 x} \varrho\, {P}^{{x}}\, ,\label{GodS13}
\end{equation}
where we define
\begin{equation}
\varrho\equiv\frac{\sqrt{2}\, J}{c^2 m}\, . \label{GodkDef}
\end{equation}

With these results and definitions, we can replace \eqref{GodS13} in
\eqref{GodTul2}, and use the constant \eqref{consEGodel} to obtain
\begin{equation}
{S}^{{xy}}=\frac{E \varrho e^{-w_0 x}}{2 c \varrho w_0+4}\, .\label{GodS23}
\end{equation}
Also, combining \eqref{GodTul1}, \eqref{GodTul2} and \eqref{GodS13} we get
\begin{equation}
\frac{2 e^{w_0 x} {S}^{{xy}}-2 c {S}^{{tx}}}{\varrho}=c e^{w_0 x} {P}^{{y}}\, .\label{GodInter}
\end{equation}
We solve for the last component of the spin using Eqs.
\eqref{GodInter}, \eqref{GodS23} and the constants $C_{3}$ and $E$
\begin{equation}
{S}^{{tx}}=\frac{\varrho e^{-w_0 x} \left(E \varrho w_0 e^{w_0 x}-C_3 \left(c \varrho w_0+2\right)\right)}{2 \left(c^2 \varrho^2 w_0^2-4\right)}\, ,\label{GodS12}
\end{equation}
and replacing ${S}^{{tx}}$ and ${S}^{{xy}}$ in \eqref{GodInter} we find
\begin{equation}
{P}^{{y}}=\frac{e^{-2 w_0 x} \left(c C_3 \left(c \varrho w_0+2\right)-2 E e^{w_0 x}\right)}{c \left(c^2 \varrho^2 w_0^2-4\right)}\, .\label{GodPy}
\end{equation}
To solve for $P^{t}$ we use Eq.~\eqref{GodTul2} along with the previous results to get
\begin{equation}
{P}^{{t}}=\frac{E-2 c C_3 e^{-w_0 x}}{2 c^2 \left(c \varrho w_0-2\right)}\, .\label{GodPt}
\end{equation}
Lastly, for $P^{x}$ we can use \eqref{GodM2}
\begin{eqnarray}
{P}^{{x}}&=&\pm \frac{e^{-w_0 x}}{2 c \left(\lambda^2-4\right)} \left[e^{2 w_0 x} \left(E^2 \left(\lambda ^2-4 \lambda -4\right)-4 c^4 \left(\lambda ^2-4\right)^2 m^2\right)\right.\nonumber\\
&&\qquad\qquad\qquad\left.-2 c^2 C_3^2 (\lambda +2)^2+8 c C_3 E (\lambda +2) e^{w_0 x}\right]^{1/2}\, ,\label{GodPx}
\end{eqnarray}
where we have defined the dimensionless constant $\lambda\equiv c
\varrho w_{0}= (\sqrt{2} J w_{0})/(c m)$. Now, our objective is to
solve for $\dot{y}$ and $\dot{x}$. Let us start by observing that
performing a time derivative of Eq.~\eqref{GodS13} yields
\begin{equation}
\dot{{S}^{{ty}}}= \varrho e^{-w_0 x} \dot{{P}^{{x}}}-w_0 \dot{x} {S}^{{ty}}\, .\label{GodDotS13}
\end{equation}
Replacing $\dot{{S}^{{ty}}}$ in \eqref{GodS2} and combining this
result with \eqref{GodP2} we get, after some algebra, an expression
for $\dot y$ In the same way, to get $\dot{x}$, we use
Eq.~\eqref{GodP3} along with the results for the spin and momentum
components. Both results are
\begin{eqnarray}
\dot{y}&=&\frac{2 (\lambda -2) \dot{{P}^{{x}}}}{C_3 w_0}\, ,\label{GodDotYPre}\\
\dot{x}&=&-\frac{2 c^2 (\lambda -2) (\lambda +2)^2 e^{w_0 x} {P}^{{x}}}{2 c C_3 (\lambda +2)^2+E ((\lambda -4) \lambda -4) e^{w_0 x}}\, .\label{GodDotXPre}
\end{eqnarray}
We now solve for $P^x$. Using \eqref{GodJ2}, \eqref{GodTul1},
\eqref{GodTul2} and \eqref{GodM2} we get
\begin{equation}
{P}^{{x}}=\pm \left(c \sqrt{\frac{2 \left({S}^{{tx}}\right)^2}{c^2
\varrho^2}-m^2-\left({P}^{{t}}\right)^2}\right)\, .
\end{equation}
Differentiate with respect to time and get
\begin{equation}
\dot{{P}^{{x}}}=\frac{c^2}{{P}^{{x}}}{ \left(\frac{2 \dot{{S}^{{tx}}} {S}^{{tx}}}{c^2 \varrho^2}-\dot{{P}^{{t}}} {P}^{{t}}\right)}\, ,\label{GodDotP2}
\end{equation}
while the time derivatives we need are
\begin{equation}
\dot{{S}^{{tx}}}=\frac{C_3 \lambda  \dot{x} e^{-w_0 x}}{2 c (\lambda -2)} \, ,\qquad \dot{{P}^{{t}}}=\frac{C_3 w_0 \dot{x} e^{-w_0 x}}{c (\lambda -2)}\, . \label{GodDers}
\end{equation}
Thus, finally, using \eqref{GodDers}, \eqref{GodDotP2} and \eqref{GodDotXPre} into \eqref{GodDotYPre} we can show that
\begin{equation}
\dot{y}=-\frac{2 c (\lambda +2) e^{-w_0 x} \left(c C_3 (\lambda +2)-2 E e^{w_0 x}\right)}{2 c C_3 (\lambda +2)^2+E \left(\lambda ^2-4 \lambda -4\right) e^{w_0 x}}\, .\label{GodDotY}
\end{equation}

In the same spirit than the previous section we can evaluate the
line element of the trajectories of the top in G\"odel spacetimes to
inquire about the nature of its orbits. With all the previous
results, and using the metric \eqref{godelmetricg}, we can find the
line element of the spin particle
\begin{equation}
\frac{ds^2}{c^2dt^2}=\frac{\dot{y}^2 e^{2 w_0 x}-2\dot{x}^2+4c \dot{y} e^{w_0 x}}{2 c^2}+1=
\frac{4 c^2 e^{2 w_0 x} \left({4 E^2 \lambda ^2}/{c^2}+c^2 (\lambda -2)^2 (\lambda +2)^4 m^2\right)}{\left(2 c C_3 (\lambda +2)^2+E \left(\lambda ^2-4 \lambda -4\right) e^{w_0 x}\right)^2}\, .
\end{equation}
As it can be readily seen, the line element is always timelike.
Thus, massive spinning particles moving in the plane $z=0$ of
G\"odel spacetimes never follow lightlike or spacelike trajectories.

\section{Exact solution for general Schwarzschild-like spacetimes}

In the preceding sections we have obtained the exact solutions of
the motion of tops in cosmological scenarios. We showed that it
could be possible for the top to follow trajectories which may be
(at least partially) described by lightlike or spacelike line
elements. Now we will study the top's dynamics in a general
spacetime which is the generalization of the results for the
Schwarzschild metric studied in Ref.~\cite{gane}.

Consider a general Schwarzschild-like metric in spherical coordinates, given by the
line element
\begin{equation}\label{lineelementgeneric}
ds^2=g_{\mu\nu}d x^\mu d x^\nu=g(r) dt^2-\frac{c^2}{g(r)} dr^2- r^2 (d\theta^2+\sin^2\theta d\phi^2)\, ,
\end{equation}
where now $g(r)\equiv g$ is a generic function with radial
dependence only. Examples of this kind of metrics includes the
Reissner-Nordstrom-(Anti)de Sitter case.

As in previous sections, we start listing the Killing vectors of this metric
\begin{equation}
\begin{aligned}
 &\xi^{0}_\mu=(g,0,0,0) \, ,\\
 &\xi^{1}_\mu=(0,0,0,-r^2 \sin^2\theta)\, , \\
 &\xi^{2}_\mu=(0,0,r^2 \sin\phi,r^2 \cos\theta \cos\phi \sin\theta)\, ,\\
 &\xi^{3}_\mu=(0,0,-r^2 \cos \phi,r^2 \cos\theta \sin\theta \sin\phi) \, .
 \end{aligned}
\end{equation}
We can use these Killing vectors to find the constants of motion. In
this case, as the metric is time-independent, we will be able to
find the four constants of motion in a straightforward manner. To
study the top's trajectories, in a way similar to the one used for
the FRW metric, let us consider its motion in the plane
$\theta=\pi/2$. First, we can find the constants of motion
\eqref{mass} and \eqref{spin} as
\begin{eqnarray}
m^2 c^2=-\frac{c^2 \left(P^{r}\right)^2}{g}+g \left(P^{t}\right)^2-r^2 \left(P^{\phi }\right)^2\, , \label{massgeneric}\\
J^2=-c^2  (S^{tr})^2+\frac{c^2 r^2}{g} (S^{r \phi})^2-g r^2 (S^{t \phi })^2\, .\label{Jgeneric}
\end{eqnarray}
And using the Killing vectors, we can find the constants of motion
\begin{eqnarray}
E=g {P}^{{t}}-\frac{g' {S}^{{tr}}}{2}\, , \label{constantEjgeneric1}\\
j=-r \left(r {P}^{\phi }+{S}^{{r\phi}}\right)\, . \label{constantEjgeneric2}
\end{eqnarray}
For the case of the general Schwarzschild-like metric, the constant
$E$ now corresponds to the energy of the top, whereas $j$ is the
conserved angular momentum orthogonal to the plane of the motion. On
the other hand, the Tulczyjew contraints \eqref{constraint} for this
case read
\begin{eqnarray}
 -\frac{{P}^{{r}} {S}^{{tr}} c^2}{g}-r^2 {P}^{\phi } {S}^{{t\phi}}=0\, , \label{Tul9generic}\\
 r^2 {P}^{\phi} {S}^{{r\phi}}+g {P}^{{t}}{S}^{{tr}}=0\, , \label{Tul10generic}\\
 \frac{c^2 {P}^{{r}} {S}^{{r\phi}}}{g}-g {P}^{{t}} {S}^{{t\phi}}=0\, . \label{Tul11generic}
\end{eqnarray}

Now, just for matter of completeness, we will show the momentum
equations \eqref{momentummotion} in their explicit form for this
metric
\begin{eqnarray}
 \dot{{P}^{{t}}}+\frac{{P}^{{r}} g'}{2 g}+\frac{\dot{r} {P}^{{t}} g'}{2 g}-\frac{r \dot{\phi } {S}^{{t\phi}} g'}{2 c^2}-\frac{\dot{r} {S}^{{tr}} g''}{2 g}=0\, , \\
 \dot{{P}^{{r}}}+\frac{g {P}^{{t}} g'}{2 c^2}-\frac{g r \dot{\phi }{P}^{\phi }}{c^2}-\frac{r \dot{\phi } {S}^{{r\phi}} g'}{2 c^2}-\frac{g {S}^{{tr}} g''}{2 c^2}-\frac{\dot{r} {P}^{{r}} g'}{2 g} =0\, ,\\
 \dot{{P}^{\phi }}+\frac{\dot{\phi } {P}^{{r}}}{r}+\frac{\dot{r} {P}^{\phi }}{r}+\frac{\dot{r} {S}^{{r\phi}} g'}{2 g r}-\frac{g {S}^{{t\phi}} g'}{2 c^2 r}=0\, ,
\end{eqnarray}
while the spin equations \eqref{spinmotion} are
\begin{eqnarray}
 \dot{{S}^{{tr}}}+{P}^{{r}}-\dot{r} {P}^{{t}}-\frac{g r \dot{\phi } {S}^{{t\phi}}}{c^2}=0\, , \label{firstspingeneric}\\
 \dot{{S}^{{t\phi}}}-\dot{\phi } {P}^{{t}}+{P}^{\phi}+\frac{\dot{\phi }{S}^{{tr}}}{r}+\frac{\dot{r} {S}^{{t\phi}}}{r}+\frac{{S}^{{r\phi}} g'}{2 g}+\frac{\dot{r} {S}^{{t\phi}} g'}{2 g} =0\, ,\\
 \dot{{S}^{{r\phi}}}-\dot{\phi } {P}^{{r}}+\dot{r} {P}^{\phi }+\frac{\dot{r} {S}^{{r\phi}}}{r}+\frac{g {S}^{{t\phi}} g'}{2 c^2}-\frac{\dot{r} {S}^{{r\phi}} g'}{2 g}=0\, .\label{lastspingeneric}
 \end{eqnarray}

 It is important to mention that the
 constants of motion \eqref{constantEjgeneric1} and \eqref{constantEjgeneric2}
 can be derived from the above set of equations.
 With this full set of equations \eqref{massgeneric}-\eqref{lastspingeneric}
 we can solve completely for the motion of  the top.
 Using Eqs.~\eqref{Tul9generic} and \eqref{Tul10generic} we get
 \begin{equation}
 \frac{r^4 \left({P}^{\phi }\right)^2 \left({S}^{{r\phi}}\right)^2}{g}-\frac{g r^4 \left({P}^\phi \right)^2 \left({S}^{{t\phi}}\right)^2}{c^2}=g \left({P}^{{t}}\right)^2 \left({S}^{{tr}}\right)^2-\frac{c^2 \left({P}^{{r}}\right)^2 \left({S}^{{tr}}\right)^2}{g}\, ,
  \end{equation}
Thereby, using \eqref{massgeneric} and \eqref{Jgeneric} we can solve for ${S}^{{tr}}$ finding that
  \begin{equation}\label{Strgenericsol}
  {S}^{{tr}}=\pm \frac{J r}{c^2 m}{P}^{\phi }\, .
  \end{equation}

New relations among the constants of motion can be found using the
constants \eqref{constantEjgeneric1}, \eqref{constantEjgeneric2} and
Eq.~\eqref{Tul10generic}. They produce the relation
  \begin{equation}
  E {S}^{{tr}}-j r {P}^{\phi }=r^3 \left({P}^{\phi }\right)^2-\frac{1}{2} g' \left({S}^{{tr}}\right)^2\, ,
  \end{equation}
which can be used to find the solution of $P_\phi$. Using
Eq.~\eqref{Strgenericsol}, after some algebra, we can find
  \begin{equation}\label{Pphisolgeneric}
  P_{\phi}=\frac{-j\pm {E J}/(mc^2)}{\eta -1}\, ,
  \end{equation}
where we introduce the notation
  $$\eta\equiv\frac{J^2 g'}{2 c^4 m^2 r}\, ,$$
  in a similar fashion, the parameter defined in Ref.~\cite{gane}.
Now we are able to solve for $P_{t}$. Using Eqs.~\eqref{constantEjgeneric1}, \eqref{Strgenericsol} and \eqref{Pphisolgeneric}, we find that
  \begin{equation}
  P_{t}=\frac{E\mp {j J g'}/(2 mc^2 r)}{1-\eta }\, ,
  \end{equation}
which lead us to easily solve for $P^{r}$ by using \eqref{mass}
  \begin{equation}
  P^{r}=\pm \frac{1}{c} \sqrt{P_t^2-g \left(c^2 m^2+\frac{P_{\phi }^2}{r^2}\right)}
  \end{equation}

  We would like to solve for $\dot{r}$ and $\dot{\phi}$. With this purpose in mind, we have to find first
  the components of the spin in terms of the momenta.
  By using constraints \eqref{Tul9generic} and \eqref{Tul10generic} along with Eq.~\eqref{Strgenericsol}
  we can show that the other components of the spin may be expressed as
  \begin{eqnarray}\label{otherspincompogheneric}
  {S}^{{t\phi}}=\mp\frac{J{P}^{{r}}}{m g r}\, ,\qquad {S}^{{r\phi}}=\mp\frac{ Jg {P}^{{t}}}{mc^2r}\, .
  \end{eqnarray}

To find $\dot{r}$, we multiply Eq.~\eqref{Tul11generic} by $\pm {Jr}/(c^2 m)$
and then we subtract Eq.~\eqref{firstspingeneric}. With the help of the spin components \eqref{otherspincompogheneric} we get
  \begin{equation}
  \frac{g' {P}^{{r}} J^2}{2 m^2c^4 r}-\frac{\dot{r} g' {P}^{{t}} J^2}{2m^2c^4 r}\pm\left(r \dot{{P}^{\phi }}+\dot{r} {P}^{\phi }\right) \frac{J}{c^2 m}-{P}^{{r}}+\dot{r} {P}^{{t}}-\dot{{S}^{{tr}}}=0\, ,
  \end{equation}
which, using \eqref{Strgenericsol}, allow us to find the solution
  \begin{equation}
  \dot{r}=\frac{{P}^{{r}}}{{P}^{{t}}}=\frac{g P^r}{P_t}\, .
  \end{equation}
  To solve for $\dot{\phi}$ we replace $\dot{r}$ from the previous
  equation in Eq.~\eqref{firstspingeneric}, as well as, we have to
  use \eqref{Strgenericsol} and Eq.~\eqref{Pphisolgeneric} (to find $\dot{P^{\phi}}$).
  After some algebra we get
  \begin{equation}
  \dot{\phi}=\frac{g P_{\phi } \left(g'-\eta  r g''\right)}{(\eta -1) r^2 g' P_t}\, .
  \end{equation}

  When the general case is reduced to
  the Schwarzschild spacetime, where $g(r)=c^2(1-2r_0/r)$ and $r_0$
  is half of the Schwarzschild radius, the
  dynamics of the top's motion described in \cite{gane} is
  recovered.

  Now we can seek for the line element \eqref{lineelementgeneric} for this metric. This is written in the plane $\theta=\pi/2$ as
  \begin{equation}
  \frac{ds^2}{c^2dt^2}=\frac{g}{c^2}-\frac{\dot{r}^2}{g}-\frac{r^2 \dot{\phi}^2}{c^2}=\frac{m^2\left(1-\Lambda\right)}{\left(P^t\right)^2}\, ,
  \end{equation}
  where we have defined the parameter
  \begin{equation}\label{lambdaGSmetricSuperl}
  \Lambda \equiv\frac{P_{\phi }^2}{c^2 m^2 r^2} \left[\frac{\left({\eta  r g''}/{g'}-1\right)^2}{(\eta -1)^2}-1\right]=\frac{\left(-j\pm {E J}/({c^2 m})\right)^2}{c^2 (\eta -1)^2 m^2 r^2}\left[\frac{\left({\eta  r g''}/{g'}-1\right)^2}{(\eta -1)^2}-1\right]\, .
  \end{equation}

As in the previous sections, according to the sign of $ds^2$ the
solution may describe timelike, lightlike or spacelike orbits,
depending on the value of $\Lambda$. We can see that here,
analoguosly to the case of the Schwarzschild metric \cite{gane},
$\Lambda$ can take different values depending on the constants of
motion of the top, such as its mass, its energy $E$, its angular
momentum $j$, and its spin. If $\Lambda<1$, the top follows timelike
trajectories. Instead if $\Lambda=1$ or $\Lambda>1$, the spinning
particle follows lightlike or spacelike trajectories (at least
partially).

\subsection{Reissner-Nordstrom-(Anti)de Sitter metric}

To evaluate the previous calculations in explicitly scenario, let us calculate $\Lambda$, from \eqref{lambdaGSmetricSuperl},  for the Reissner-Nordstrom-(Anti)de Sitter.  This metric is given by
\begin{equation}
g(r)=c^2 \left(1-\frac{2 G M}{c^2 r}+\frac{\kappa G Q^2}{c^4 r^2}-\frac{\lambda  r^2}{3}\right)\, ,
\end{equation}
where $G$ is the gravitational constant, $\kappa$ is Coulomb's constant, $Q$ is the charge of the black hole, $M$ its mass and $\lambda$ is the cosmological constant.

For this case, the parameter \eqref{lambdaGSmetricSuperl} becomes
\begin{equation}
\Lambda=\frac{27 c^6 G J^2 r^6 \left(3 c^2 M r-4 \kappa Q^2\right) \left(c^2 j m+E J\right)^2 \left(6 c^6 m^2 r^4+2 c^4 J^2 \lambda  r^4+3 c^2 G J^2 M r-6 \kappa G J^2 Q^2\right)}{\left(3 c^6 m^2 r^4+c^4 J^2 \lambda  r^4-3 c^2 G J^2 M r+3 \kappa G J^2 Q^2\right)^4}\, .\label{lambdaRN}
\end{equation}
We have found that searching for parameter combinations that produce $\Lambda\geq 1$ is a very difficult task when selecting known values for different particles. For example, trying to make an electron accelerate to the speed of light is practically impossible on an extremely large range of central object masses (from Earth like to extreme black holes) under even the most extreme scenarios for values of $E$ and $j$.
Another interesting result is that the cosmological constant has no effect on the line element when there is no central object present ($M=Q=0$). This can easily be seen from Eq.~\eqref{lambdaRN}.

\section{Exact solution for static spherically symmetric conformally flat spacetimes}

As well as in the other metrics studied in this work, the motion of
a top in a conformally flat spacetime with spherical symmetry may be
solved exactly. The conformally spherical line element is
\begin{equation}\label{metriConf}
ds^2=g_{\mu\nu}d x^\mu d x^\nu\, ,\qquad  g_{\mu\nu}=\Omega^2\eta_{\mu\nu}\, ,
\end{equation}
where $\Omega\equiv\Omega(r)$ is the spherical symmetric conformal
factor, and $\eta_{\mu\nu}$ as the flat spacetime metric written in
spherical coordinates, such that $\eta_{\mu\nu}d x^\mu d x^\nu=c^2
dt^2-dr^2-r^2 (d\theta^2+\sin^2\theta d\phi^2)$.

Because the metric is time-independent, we can find a general
expression for the conserved energy using the general conserved
quantity \eqref{ckilling}. We get the conserved energy $E$ for the
motion of the top
\begin{equation}\label{energy1}
  E=P_t-c^2\Omega\Omega'  S^{tr}\, ,
  \end{equation}
associated to the Killing vector $\xi_\mu^1=(c^2\Omega^2,0,0,0)$.
Similarly, the three components of the conserved vectorial angular
momentum  of the top can be found in a similar fashion. Without loss
of generality, we again restrict ourselves to motion in the
equatorial plane $\theta=\pi/2$, as in the previous sections. The
conserved angular momentum component orthogonal to the orbital plane
is
\begin{equation}\label{angularmom1}
  j=-\Omega\left(\Omega+r\Omega'\right) r S^{\phi r}-P_\phi\, ,
\end{equation}
which can be checked from \eqref{ckilling} using the Killing vector $\xi_\mu^2=(0,0,0,-r^2\Omega^2)$.

On the other hand, the mass and spin conservation laws give rise to new relations.
From \eqref{mass} we find that
\begin{equation}\label{condmm}
  \frac{r^2}{c^2}P_t^2-r^2 P_r^2-P_\phi^2=r^2\Omega^2 m^2 c^2\, ,
\end{equation}
whereas from \eqref{spin} we have
\begin{equation}\label{condss}
  J^2=\Omega^4\frac{(S^{\phi r})^2}{P_t^2}\left(r^2P_t^2-c^2 P_\phi^2-c^2 r^2P_r^2\right)\, ,
\end{equation}
where use has been made of the constraints \eqref{constraint}
\begin{equation}\label{constspinmom}
S^{tr}=-\frac{S^{\phi r}P_\phi}{P_t}\, ,\qquad
S^{t\phi}=\frac{S^{\phi r}P_r}{P_t}\, .
\end{equation}
We notice that Eqs.~\eqref{condmm} and \eqref{condss} can be used to get the
condition
\begin{equation}\label{sphir2}
S^{\phi r}=\pm \frac{J P_t}{m r \Omega^3 c^2}\, .
\end{equation}
This condition allows us to rewrite the conserved energy and angular
momentum in terms of momenta. Using \eqref{constspinmom} and
\eqref{sphir2}, we can write the energy \eqref{energy1} and the
angular momentum \eqref{angularmom1} as
\begin{equation}\label{energy2}
  E=P_t\pm\frac{J\Omega'}{r m\Omega^2} P_\phi\, ,
\end{equation}
\begin{equation}\label{angularmom2}
  j=\mp\frac{J\left(\Omega+r\Omega'\right)}{\Omega^2 m c^2}P_t-P_\phi\, .
\end{equation}

Notice that two solutions for the energy and the angular momentum
have emerged now. Their origin are the two possible solutions for
$S^{\phi r}$ of the condition \eqref{sphir2} related to the fact
that the spin vector may be parallel or antiparallel to the angular
momentum vector.

The previous relations can be solved for $P_t$ and $P_\phi$. It is
again convenient to define the dimensionless auxiliary function
\begin{equation}
 \eta\equiv\eta(r)=\frac{\Omega'\left(\Omega+r\Omega'\right)J^2}{r\Omega^4 m^2 c^2}\, ,
\label{eta}
\end{equation}
in order to find the top's momentum from \eqref{energy2} and \eqref{angularmom2}
\begin{equation}
 P_\phi=\frac{1}{1-\eta}\left(-j\mp \frac{E J(\Omega+r\Omega')}{m c^2\Omega^2}\right)\, ,
\label{pfi}\end{equation}
\begin{equation}
 P_t=\frac{1}{1-\eta}\left(E\pm \frac{j J \Omega'}{m r \Omega^2}\right)\, .
\label{pt}\end{equation}
Lastly, from the condition \eqref{mass}, we get
\begin{equation}
 P_r=\pm \left[\frac{P_t^2}{c^2}-\frac{P_\phi^2}{r^2}-m^2c^2\Omega^2\right]^{1/2}\, .
\label{pr1}\end{equation}

Once we find the momenta in terms of the conserved quantities, it is
possible to solve the system for the velocities $\dot r$ and
$\dot\phi$. With the help of \eqref{constspinmom}, the equations of
motion \eqref{spinmotion} for $S^{tr}$ and $S^{t\phi}$ become
\begin{eqnarray}\label{}
P^t \dot r-P^r&=&\frac{S^{\phi r}P_\phi}{P_t^2}\frac{D P_t}{D\lambda}-\frac{D S^{\phi r}}{D\lambda}\frac{P_\phi}{P_t}-\frac{S^{\phi r}}{P_t}\frac{D P_\phi}{D\lambda}\, ,\\
P^t \dot \phi-P^\phi&=&-\frac{S^{\phi r}P_r}{P_t^2}\frac{D P_t}{D\lambda}+\frac{D S^{\phi r}}{D\lambda}\frac{P_r}{P_t}+\frac{S^{\phi r}}{P_t}\frac{D P_r}{D\lambda}\, ,
\end{eqnarray}
where this time we have not written explicitly the covariant
derivatives. Using \eqref{momentummotion}, these equations may be
solved for $\dot r$ and $\dot\phi$ in the equatorial plane to give
\begin{equation}
 \dot\phi=\frac{\zeta\, \gamma\, c^2}{r^2}\left(\frac{P_\phi}{P_t}\right)\, ,\quad
 \dot r=\zeta\, c^2 \left(\frac{P_r}{P_t}\right)\, ,\quad
\frac{d\phi}{dr}=\frac{\gamma}{r^2}\left(\frac{P_\phi}{P_r}\right)\, ,
\label{dphidr}
\end{equation}
where we have defined
\begin{equation}\label{alfa}
  \zeta=(\eta-1)\left[\eta+1-\frac{J^2}{\Omega^3 m^2 c^2 r}\left(2\Omega'+r\Omega''\right)\right]^{-1}\, ,
\end{equation}
and
\begin{equation}\label{gama}
  \gamma=(1-\eta)^{-1}\left[1+\frac{J^2}{\Omega^4 m^2 c^2}\left((\Omega')^2-\Omega\Omega''\right)\right]\, .
\end{equation}

Thereby, the motion of the spinning massive particle has been solved
exactly, remembering that $J$ may be identified with the top's spin.
It is worth noting that the  preceding expressions coincide with the
usual results for geodesic motion when the spin is neglected, $J=0$,
being $\eta=0$, $P_\phi=-j$, $P_t=E$, and
$P_r^2=E^2/c^2-j^2/r^2-m^2c^2\Omega^2$. Thus, the velocities are
reduced to $\dot r=-c^2 P_r/E$ and $\dot\phi=c^2j/(r^2 E)$.

Again, another interesting aspect of the motion of the top is the
evaluation of its interval \eqref{metriConf}. This becomes
\begin{eqnarray}
\frac{ds^2}{c^2 dt^2}
 &=&\Omega^2\left(1-{\zeta^2}\right)+\frac{m^2 c^4 \zeta^2\, \Omega^4}{(P_t)^2}(1-\Lambda)\, ,
\label{dsdt}
\end{eqnarray}
where we have introduced the parameter
\begin{equation}
 \Lambda=\frac{(P_\phi)^2 \left(\gamma^2-1\right)}{\Omega^2 m^2 r^2 c^2}\, .
 \label{lambdasuperluminal}
\end{equation}

 From \eqref{dsdt} we realize that, even for massive particles,
 there could be some initial conditions such that $ds^2\leq0$,
 at least in part of the top's trajectories.
The contribution of
$\Lambda$  depends on the conformal factor and also depends
strongly on the value of the mass, being important for small mass values.
When the spin is neglected, then $ds^2=c^2dt^2\Omega^4(m^2 c^4/E^2)>0$, and
the top always travels in timelike trajectories.

To show this behavior with explicit conformal metric,
we will study different scenarios where the conformal metric is relevant.

\subsection{Conformally flat spacetime for static perfect fluid and radiation}

As an example, we study the motion of a test top in a conformally
spherically flat universe filled with a static perfect fluid and
radiation. We will focus our attention in four different cases:
radiation only, matter-dominated, radiation-dominated and
inflation-dominated universes. The universe will be filled with a
static fluid, with  energy density $\epsilon$ and  pressure $p$, and
a electromagnetic radiation field $F_{\mu\nu}$ which has origin in a
electrostatic potential, i.e., its only non-vanishing component is
$F_{0r}$.

Assuming that the fluid pressure $p$ is proportional to the fluid
energy density $\epsilon$, the Einstein equations for the conformal
metric \eqref{metriConf} can be solved exactly. A general conformal
factor can be found  \cite{banerjee}
\begin{equation}\label{Congeneral}
\Omega(r)=Q_1(r^{-2/(1+3\alpha)}-Q_2 \beta)^{(1+3\alpha)/2}\, ,
\end{equation}
where $Q_1$ and $Q_2$ are constants, $\alpha=p/\epsilon$ is the
constant ratio between pressure and the energy density,  and
$\beta=(1+3\alpha)^{(3+3\alpha)/(1+3\alpha)}$. Also, the energy
density has the form \cite{banerjee}
\begin{equation}\label{energyGener}
  \epsilon=\frac{3c^2\beta Q_2}{4\pi G Q_1^2(1+3\alpha)}{r^{-\frac{4+6\alpha}{1+3\alpha}}}\left(r^{-2/(1+3\alpha)}-Q_2\beta \right)^{-3-3\alpha}\, ,
\end{equation}
where $G$ is the gravitational constant. Notice that we must require
$Q_2\geq0$ in order to have a positive semidefinite energy density.
The radiation field can be found to be
\begin{eqnarray}\label{radiationGener}
  F^2&=&\frac{-c^2 r^{\frac{-6-6\alpha}{1+3\alpha}}}{G Q_1^2(1+3\alpha)}\left(r^{\frac{-2}{1+3\alpha}}-Q_2\beta \right)^{-3-3\alpha}\left[2+6\alpha-4(2+3\alpha)\beta Q_2 r^{\frac{2}{1+3\alpha}} \right]\, ,
\end{eqnarray}
where $F^2\equiv F_{\mu\nu} F^{\mu\nu}=2F_{0r} F^{0 r}=-2(F_{0r})^2/(\Omega^4c^2)$.
Because $Q_2\geq0$, this solution could have an intrinsic singularity in the metric \cite{banerjee}.

\subsubsection{Uniform electromagnetic radiation universe}
\label{radU}

If $Q_2=0$, the conformal factor will be simply
\begin{equation}\label{}
\Omega=\frac{Q}{r}\, .
\end{equation}
This conformal metric describing this kind of  universe is known as
the Bertotti-Robinson solution \cite{berto,robin,stefani}. It is
important to notice from \eqref{energyGener} that the energy density
and the pressure vanish. Therefore, there is no fluid. On the other
hand, from \eqref{radiationGener}, we get that $F^2=-{2c^2}/(G Q^2)$
is constant. Thus, this universe is filled only with a static
uniform electromagnetic radiation field.

The motion of a top in this universe can be studied using the
Bertotti-Robinson metric. From \eqref{eta} we find that $\eta\equiv
0$. This fact simplifies the previous expressions, finding from
\eqref{pfi}, \eqref{pt} and \eqref{gama}, that
\begin{equation}\label{Ptradi}
P_\phi=-j\, ,\qquad P_t=E\pm\frac{j J}{m Q r}\, ,\qquad \gamma=1-\frac{J^2}{m^2 c^2 Q^2}\, ,
\end{equation}
respectively.  Also, from \eqref{alfa} we get $\zeta=-1$. This
implies that the nature of the interval $ds^2$ of the top particle
\eqref{dsdt} is controlled by $\Lambda$ only. From
\eqref{lambdasuperluminal} we find that
\begin{equation}\label{}
  \Lambda=\frac{j^2 J^2}{m^6 c^6 Q^6}\left(J^2-2 m^2 c^2 Q^2\right)\, .
\end{equation}
With these quantities we can evaluate the velocities \eqref{dphidr},
and thus, the motion of the spinning particle is completely
described.

Lastly, notice that $\Lambda$ can be positive,  negative or null
depending on the properties of the top and of the spacetime through
$Q$. This determines the sign of the interval $ds^2$, which it is
\begin{equation}\label{expression1}
ds^2=\frac{dt^2}{r^4 {P_t}^2m^4 Q^2}\left(2m^2 c^2Q^2 j^2 J^2+ m^6 c^6 Q^6-j^2 J^4\right)\, ,
\end{equation}
where $P_t$ is given in \eqref{Ptradi}. If the previous expression
for $ds^2$ is positive, the spinning massive particle follows
timelike orbits. Otherwise, the particle can travel in lightlike or
spacelike orbits. Interesting enough is that this condition depends
only on the properties of the particle $j$, $J$ and $m$, and not on
the top's path. Thus, depending on the top's properties, the
previous desciption is for the motion of bradyons, luxons or
tachyons \cite{sudar}.

\subsubsection{Matter-dominated universe}

If we choose $\alpha=0$, then we describe a matter-dominated
universe \cite{ryden}. In this case, the energy density is dominant
over the pressure, and the fluid is composed by non-relativistic
matter (cold dust, $p=0$). For this case, the conformal factor will
be given by
\begin{equation}\label{}
\Omega=Q_1\left(\frac{1}{r^{2}}-Q_2\right)^{1/2}\, ,
\end{equation}
 making the evaluation of the $\eta$ function a straightforward calculation
\begin{equation}\label{etadust}
  \eta=\frac{J^2 Q_1^4 Q_2}{r^4m^2c^2 \Omega^6}>0\, .
\end{equation}
This allows us to write the other auxiliary functions in terms of
$\eta$. We find that $\zeta=(\eta-1)/(2\eta+1)$ and
$\gamma=[1+3\eta-\eta/(r^2Q_2)]/(1-\eta)$. In this case, the momenta
\eqref{pfi} and \eqref{pt} take the form
\begin{equation}\label{Ppsdust}
  P_\phi=\frac{1}{1-\eta}\left(-j\pm\frac{EJQ_1^2Q_2}{mc^2\Omega^3}\right)\, ,\qquad   P_t=\frac{1}{1-\eta}\left(E\mp\frac{jJQ_1^2}{mr^4\Omega^3}\right)\, .
\end{equation}
With these values, the evaluation of $P_r$ or of $F^2$ are could now
be done using \eqref{pr1} and \eqref{radiationGener}.

Along with this results, it is possible to obtain an expression for
the interval \eqref{dsdt} in this spacetime. As in the previous
case, the spin introduces enough freedom to determine the sign of
$ds^2$.
 It can be shown from \eqref{dsdt} that
\begin{eqnarray}\label{expression2}
\frac{ds^2}{c^2dt^2}&=&\frac{\Omega^2}{(\eta+1-2\eta^2)^2{P_t}^2}\left[ 3\eta(2+\eta)\left(1-\eta\right)^2{P_t}^2+m^2 c^4\Omega^2(1-\eta)^4\right.\nonumber\\
&&\left.-\left(1-\eta\right)^2{P_\phi}^2\left(\frac{8\eta(1+\eta)c^2}{r^2}+\left(\frac{J^2 Q_1^4}{\Omega^{6} c\, m^2  r^7}\right)^2-\frac{(2+6\eta)J^2 Q_1^4}{\Omega^6 m^2 r^8}\right)\right]\, ,
\end{eqnarray}
written in terms of $\eta$ given in \eqref{etadust} and the momenta
\eqref{Ppsdust}. If this expression for $ds^2$ is null, the motion
will be lightlike, and it will be spacelike if the right-hand side
of \eqref{expression2} is negative. Otherwise, the trajectory will
be timelike. In addition to its dependence on the top's properties,
the above expression for $ds^2$ also depends on $r$, implying that
the particle can achieve velocities larger than speed of light in
certain regions of its path. A rigorous evaluation of this condition
requires the knowledge of the top's energy $E$, mass $m$ and angular
momentum $j$, and the constants $Q_1$ and $Q_2$ of the metric.

\subsubsection{Radiation-dominated universe}

The radiation-dominated phase of the universe (very early universe)
can be studied for $\alpha=1/3$ \cite{ryden}, implying an equation
of state for ultra-relativistic matter. In this case, the conformal
factor \eqref{Congeneral} becomes
\begin{equation}\label{}
\Omega=Q_1\left(\frac{1}{r}-4Q_2\right)\, ,
\end{equation}
which produces now that
\begin{equation}\label{etaRad}
  \eta=\frac{4J^2 Q_1^2 Q_2}{r^3m^2c^2 \Omega^4}>0\, .
\end{equation}
Then, the auxiliary functions become $\zeta=(\eta-1)/(\eta+1)$, and
$\gamma=[1+2\eta-\eta/(4rQ_2)]/(1-\eta)$, whereas the momenta are
\begin{equation}\label{momePPradiation}
  P_\phi=\frac{1}{1-\eta}\left(-j\pm\frac{4EJQ_1Q_2}{mc^2\Omega^2}\right)\, ,\qquad   P_t=\frac{1}{1-\eta}\left(E\mp\frac{jJQ_1}{mr^3\Omega^2}\right)\, ,
\end{equation}
which can be used to find the velocities of the particle.

Again, using the previous solution an explicit expression for the
top's interval could be found. We again obtain a possibility for a
behavior different to a timelike interval. It can be shown from
\eqref{dsdt} that
\begin{eqnarray}\label{expression3}
\frac{ds^2}{c^2 dt^2}&=&\frac{\Omega^2 }{(1-\eta^2)^2{P_t}^2}
\left[4\eta\left(1-\eta\right)^2{P_t}^2+m^2 c^4\Omega^2(1-\eta)^4\right.\nonumber\\
&&\left.-\left(1-\eta\right)^2{P_\phi}^2\left(\frac{3\eta(2+\eta)c^2}{r^2}+\frac{J^4 Q_1^4}{\Omega^{8} m^4 c^2 r^{10}}-\frac{(2+4\eta)J^2 Q_1^2}{\Omega^4 m^2 r^6}\right)\right]\, ,
\end{eqnarray}
in terms of $\eta$ defined in \eqref{etaRad} and momenta
\eqref{momePPradiation}. Thus, $ds^2$ could define timelike,
lightlike or spacelike motion for the spinning massive particle, if
it is positive, null or negative, respectively. Anew, the lightlike
or the spacelike behavior of the top's interval will depend on the
distance and on its properties. For some appropriated values of the
energy, mass and angular momentum of the massive particle, the
spinning massive particle can be have different behaviors for the
velocities in some part of the trajectory.

\subsubsection{Inflation scenario}

As a final example, we perform a theoretical exercise studying a
universe filled with dark energy. In this case the universe will be
in an inflationary state with a cosmological constant. A simplest
case is to consider a fluid with the equation of state $\alpha=-1$
\cite{peeb,ryden}. Thus, the conformal factor acquires the form
\begin{equation}\label{}
\Omega=\frac{Q_1}{r-Q_2}\, ,
\end{equation}
while from \eqref{eta} we find that $\eta={Q_2 J^2}/(Q_1^2 r m^2
c^2)$. Remarkably, this metric produces huge simplifications. It
implies that $\zeta=-1$. Also, we obtain that
$\gamma=r(J^2-Q_1^2m^2c^2)/(Q_2J^2-rQ_1^2m^2c^2)$, and the momenta
are
\begin{equation}\label{}
  P_\phi=\frac{1}{1-\eta}\left(-j\pm\frac{EJQ_2}{mc^2Q_1}\right)\, ,\qquad   P_t=\frac{1}{1-\eta}\left(E\mp\frac{jJ}{mrQ_1}\right)\, .
\end{equation}

Finally, we get that the interval \eqref{dsdt} of the top particle
is  controlled only by $\Lambda$ (due to the fact that $\zeta^2=1$).
For this case, we get
\begin{eqnarray}\label{expression4}
\frac{ds^2}{c^2dt^2}&=&\frac{m^2c^2\Omega^4}{\left(Q_1^2 r\, m^2 c^2-Q_2 J^2\right)^4{P_t}^2}
\left[c^2\left(J^2Q_2-Q_1^2r\, m^2c^2\right)^4\right.\nonumber\\
&&\left.-J^2(r-Q_2)^3\left(j Q_1 m c^2\mp E J Q_2\right)^2\left(J^2(r+Q_2)-2Q_1^2r\, m^2c^2\right)\right]\, .\end{eqnarray}
The interval  $ds^2$ could be null or negative, and the motion could be lightlike or spacelike.
This condition  also depends on the distance.  So, a spinning massive particle moving in
a universe  with dark energy could reach  velocities larger than speed of light
due to its spin.

\section{Conclusions}

We have find exact solutions for the motion of a spinning massive
particle in different spacetimes. First, we study two are
cosmological models, the Friedmann--Robertson--Walker and  G\"odel
spacetimes. Later we study  a general Schwarzschild-like spacetime
and finally the static spherically symmetric conformally flat
spacetime.

The top's motion in each metric is exhibited in detail. For the case
of the FRW spacetime, we found a new conserved quantity that allow us to solve completely the problem.
For the general Schwarzschild-like spacetime the case of the
Reissner-Nordstrom-(Anti)de Sitter metric is shown, whereas for the
conformally flat spacetimes, we show different solutions for
different universes filled with a perfect fluid and/or radiation.

In any case, the solutions are written in terms of the momenta and
velocities of the spin particle. The spin strongly modifies the
dynamics of the spinning particle (as compared to the spinless
case), which modifies the momenta, the velocities, and, in
particular, the nature of its line-element for the orbital motion.
Although most of the solutions produce timelike trajectories with
$ds^2>0$, for various cases there are some specific relations
between the energy, the spin and the angular momentum that can give
rise to luminal ($ds^2=0$) or superluminal ($ds^2<0$) motion for
massive particles. These are the cases of the motion for the spin
particle in the FRW, general Schwarzschild-like and static
conformally flat spacetimes. These results generalize those found
for spinning massive particles in a Schwarzschild background
\cite{gane}. The results presented here seem to indicate that the
effects due to spin are robust.

 \begin{acknowledgements}
NZ thanks CONICyT-Chile for Funding N$^o$ 21080567. FAA thanks to CONICyT-Chile for Funding N$^o$ 79130002.
 \end{acknowledgements}

\end{document}